\documentclass[prl,twocolumn,showpacs,preprintnumbers]{revtex4-1}
\usepackage{amsmath,amssymb,latexsym}

\usepackage{graphicx}

\begin{document}

\title
{The simplification of the electron-ion many body problem.\\
$N$-representability of the pair densities obtained via a classical-map
for the electrons.}

\author
{
 M.W.C. Dharma-wardana}
\affiliation{
National Research Council of Canada, Ottawa, Canada, K1A 0R6
}
\email[Email address:\ ]{chandre.dharma-wardana@nrc-cnrc.gc.ca}

\date{\today}

\begin{abstract}
The classical map hypernetted-chain (CHNC) method for interacting
electrons  uses a kinetic energy functional in the form of a classical-fluid temperature.
Here we show that  the CHNC generated two-body densities and pair-distribution functions
(PDFs)  correspond to  $N$-representable densities. Comparisons of results from
CHNC with quantum Monte Carlo (QMC) and Path Integral Monte Carlo (PIMC) are used
to validate the CHNC results. Since the PDFs  are sufficient
 to  obtain the equation of state or linear-response properties of
 electron-ion systems, we apply  the CHNC method for fully
 classical calculations of
 electron-ion systems in the quantum regime, using hydrogen at 4000K and
350 times the solid density as an example since  QMC comparisons
 are available. We also present neutral  pseudo-atom (NPA)
 calculations which use rigorous density-functional theory (DFT) to reduces
the many nuclear problem to an effective  one-ion problem. 
 The CHNC PDFs and NPA results agree well with the ion-ion, electron-ion and
 electron-electron PDFs from  QMC, PIMC, or DFT coupled to molecular dynamics
 simulations where available.  The PDFs of a 2D electron-hole system at 5K
 are given as an example
of 2D `warm dense' matter where the electrons and the counter
particles (holes) are all in the quantum regime. Basic methods like QMC, PIMC
  or even  DFT
become prohibitive while  CHNC methods, being independent of the
number of particles or the temperature,  prove to be  easily deployable.
\end{abstract}

%
\maketitle
\section{Introduction}
\label{intro} 
The wavefunction $\Psi(\{\vec{r}_i\}, \{\sigma_i\},\{\vec{R}_j\})$ of an
 $N$-electron quantum
 system  depends on 3$N$ space coordinates $\vec{r}_i$, spin  $\sigma_i$,
 and the coordinates
 $\{\vec{R}_j\}$ specifying  the positions of the nuclei.
 In the following at first the  ions  are treated as passively providing
 an `external potential' to the electrons, so that we only have an electron system
placed in the `external potential' of the ion subsystem. Subsequently,
 the two component system of interacting electrons and nuclei are treated. 

 In the general case with electrons with a one-body density $n(\vec{r})$ and
 nuclei at a density $\rho(\vec{r})$, we have three many-body interaction
 terms in the Hamiltonian, viz., $V_{ee}, V_{ei}$, and $V_{ii}$.  Density
 functional theory (DFT) can be used to reduce each of these
 three terms to three effective onebody interactions and three corresponding
 exchange-correlation functionals. That is, a full DFT reduction of the
 electron-ion manybody problem will involve  the familiar $e$-$e$
 exchange-correlation
  functional  $F^{ee}_{xc}$ as well as two additional
 functionals $F^{ei}_{xc}$ and $F^{ii}_{xc}$. 
Since ions behave as classical particles under normal
conditions, $F^{ii}_{xc}[\rho(r)]$ is simply a correlation functional $F^{ii}_c$,
with exchange effects being normally of  no importance, even if the ions
 were fermions or bosons. However, if the postively charged particles are not
nuclei, but `holes' in semiconductor energy bands at ambient temperatures,
positrons or muons, exchange effects have to be included.

In  a complete DFT description of, say, a fluid of electrons and carbon nuclei,
the many-body problem would be reduced to a single `Kohn-Sham electron' interacting with
 a single `Kohn-Sham carbon ion' via Coulomb interactions and three coupled XC-functionals.
In effect, the total electron-ion problem can be reduced to an effective
 ``hydrogenic'' problem. Such a model is already available, mainly for extended systems
 like fluids and metallic solids, in the implementation of the neutral pseudo atom (NPA)
 model that was initially introduced into solid state physics~\cite{Ziman64NPA} using
linearly screened ions. Ions screened by a DFT-generated non-linear electron density were 
introduced by Dagens~\cite{Dagens75},  and then into plasmas and fluids~\cite{Chihara85}
 mainly using  cluster expansion models for dilute
 superposed densities of ions, but without recourse to an ion-ion correlation functional.
 The NPA was reformulated using  more systematic DFT arguments in
 Ref.~\cite{dwp82} where explicit ion-ion and electron-ion correlation functionals
were introduced.

 A further level of
 simplification, going beyond the NPA model is to replace the quantum electrons by an
 equivalent classical representation of electrons~\cite{prl1}.
 In effect, such models replace the quantum kinetic energy functional by a functional
for an  `effective
temperature' of an `equivalent' classical system. Then the PDFs can be determined
using classical molecular dynamics (MD)
simulations or 
classical integral equations rather than
those of quantum systems which involve fermion sign problems, problems of evaluation
of multi-center integrals and large basis sets. Thus, even at zero temperature, even
DFT calculations scale non-linearly with the number of particles, and rapidly become
prohibitive at finite temperatures, while classical methods remain feasible.  

 Most DFT calculations only use the {\it e-e} XC-functional and make no
 recourse to $F^{ei}_{xc}$
 and $F^{ii}_c$. This is perhaps because the construction of an XC-functional even for 
 electrons has been a major task since the inception of density functional theory.
 However, the three needed XC-functionals can be formally expressed  in terms of the
 three pair-distribution functions $g_{aa'}(\vec{r},\vec{r}\,')$ where $a$ stands
 for particle species $e$, or  $i$. This defines  fully non-local functionals most
appropriate to the given calculation.  In fact, if the three pair distribution functions 
 (PDFs) were known, {\it all} static quantum properties and thermodynamic properties
 as well  as  linear transport properties (e.g., electrical conductivities) can be
 determined without having to know the manybody wavefunction. Hence it would be a great
 simplification of the electron-ion manybody problem if there were accurate methods
 for the  direct determination of the PDFs via ``orbital-free'' methods. 

The objective
 of the   paper is to demonstrate a method where (a) it is shown that calculations
 using a classical map for electrons provide results for the three PDFs having an accuracy
 quite  competitive with those from DFT in `difficult' regimes like warm dense matter;
 (b) the PDFs obtained using the classical map hypernetted-chain technique 
satisfy the criteria for $N$-representability~\cite{Coleman63, Erdahl87}.

 $N$ representability  stipulates that any directly  determined PDF or two-body density
 must be such  that it is the result of integrating out  the space and spin coordinates
 of all but two of the
 particles from the square of a many-body wavefunction (which is unknown). Of course,
 $N$-representability of a two-body density  does not guarantee that it is the one
 with  minimum energy, but only that the PDF or two-body density 
is a reduction of an $N$ -body wavefunction.

It can perhaps be argued that current rapid developments in computer technology have
made  attempts at simplifications of the electron-ion problem less relevant. However,
even today, DFT calculations are unable to provide all three  PDFs
 $g_{a,a'}(\vec{r},\vec{r}\,')$ even for uniform systems like fluids or electron-hole
 layers, even at zero temperature. Quantum Monte Carlo (QMC) and path integral Monte
 Carlo (PIMC) methods become major  calculational projects when applied even to
 electron-proton  systems at finite $T$. In this study we present examples of
 competitively accurate  calculations using  a classical  map approach capable
 of reproducing such heavy
 computations  `on a laptop'  in minutes, without the use of {\it ad hoc} models and
using only the temperature, density and nuclear charge as inputs,
 for a wide class of  problems where these  methods apply.
 
\section{Survey of the theory}
Usually  the  ions behave `classically', while the electrons
are quantum mechanical. We seek  to represent the electrons 
even at the extreme quantum limit of $T=0$
by an `equivalent'  classical Coulomb fluid (CCF), only in 
the limited sense of having the same  pair distribution functions (PDFs)
as the electron system. 

The many-electron  wavefunction contains significantly more information than
necessary for calculating measurable properties of physical systems.
As $N$ becomes large, the solution of the many-particle Schr\"{o}dinger
or Dirac equation, or their QMC pr PIMC implementations
become numerically prohibitive. A way out is presented by the
Hohenberg-Kohn theorem of  density functional theory (DFT), which asserts
that the ground-state energy $E$ of the $N$-particle system is a functional
 of just the {\it one-body} electron density $n(\vec{r},\sigma)$~\cite{HK64,KS65,DG90}.
\begin{equation}
n(\vec{r},\sigma)=\Sigma_{i=2}^N\int d\vec{r}_2\hdots d\vec{r}_N |
\Psi(\vec{r}_1,...\vec{r}_N,\{\sigma_i\})|^2, 
\end{equation}
From now on we consider a paramagnetic electron fluid and suppress spin indices
unless we consider effects explicitly dependent on spin-resolution.

The Hohenberg-Kohn theorem is a counter-intuitive result since the many-particle
 Hamiltonian
\begin{equation}
H=T+V(\vec{r}) + \Sigma_{i<j}V_{ee}(\vec{r}_i,\vec{r}_j)
\end{equation}
explicitly contains the electron-electron Coulomb potential -- a two-body interaction. 
The many-body effects of this  interaction, as well as  corrections arising from
the kinetic energy operator $T$ acting on the many-body wavefunction $\Psi$ are
 contained in a one-body energy functional  known as the XC-functional of DFT,
 viz, $E_{xc}([n])$. Kohn and Sham modeled the XC-functional using results for
the uniform electron liquid (UEL) that we consider here.
The Hartee energy of the UEL is zero, and the  exchange energy component $E_x$ of
 the Hartree-Fock energy is known explicitly.
 In fact, it is given at arbitrary temperatures  in
 parametrized from in Ref.~\cite{PDW-XC84}.
The correlation energy $E_c$ is usually defined as including all
contributions to the total energy $E_T$ beyond the Hartree-Fock energy $E_{HF}$,
(c.f., Eq. 6.107 of Ref. ~\cite{DG90}).
\begin{equation}
\label{ecdef.eq}
E_c=E_T-E_{HF}
\end{equation}
The grouping of $E_x$ and $E_c$ together is needed
 (especially for free-electron systems
like metals and plasmas at finite-$T$) to accomodate important cancellations 
between the two terms~\cite{PDW-XC84}. At finite-$T$ we replace the internal energies $E$
in Eq.~\ref{ecdef.eq} by their Helmholtz free energies $F$. Furthermore, $F_x,F_c$
occur as the sum in  direct evaluations of XC-energies from the PDF~\cite{VashisKohn83}.
Thus $F_x,F_c$ are grouped together in the
XC-functional $F_{xc}$ whose functional derivative with respect to the
one-body density gives the Kohn-Sham one-body XC-potential.

\subsection{Kohn-Sham theory,  one-body and  two-body densities}
The Hohenberg-Kohn theory posits that the exact ground state
one-body density $n(\vec{r})$ is precisely the one which minimizes the
 ground state energy within a constrained search scheme. 
Its extension to finite-$T$~\cite{Mermin70} states that the Helmholtz free energy $F$
of the system is a functional of the one-body density, and that $F$ is a minimum for the
true density. The finite-$T$ theory is considered to be more robust than the $T=0$
theory, e.g.,  when magnetic fields are included~\cite{CapelleVig01,Garrigue19}.

It was recognized prior to DFT that the ground state
energy can be expressed entirely via the two-body density
 $n(\vec{r}_1,\vec{r}_2)$,
but a reduction to a one-body functional was not suspected.
The two-body density matrix  is obtained by integrating all but two of the space
and spin variables of the $N$-body density, i.e., $|\Psi(\{\vec{r}_i\},\{\sigma_i\})|^2$.
This is also known as the two-particle reduced density matrix (2-RDM), and
 identifies with the  PDF itself (depending on the prefactors used). The
 pair-distribution function
 $g(\vec{r}_1,\vec{r}_2)$ reduces to
 $g(r)$ for a uniform system, and  gives the probability of
finding a second particle at the radial distance $r$, with the first particle at
 the origin. 
 
 The one-body density in a system where the origin of coordinates is attached
 to one of the particles
 automatically becomes a 2-body density in the laboratory frame, and hence the PDF 
of homogeneous
systems, e.g., a uniform Coulomb fluid, can be used to display the inherent particle
 correlations in a uniform fluid. 
\begin{equation}
n(\vec{r}_1{\tiny\mbox=0,\vec{r}_2 \tiny\mbox=\vec{r}}) = \bar{n}\,g_{12}(r).\nonumber
\end{equation}
While  placing the origin on a classical particle is  possible, so identifying
a specific electron in the quantum problem is not possible. 

\subsection{The $N$-representability condition on the two body density matrix.}
\label{rdm.sec}
In 1955 Mayer proposed~\cite{Mayer55} to compute the
ground-state energy of $N$-electron systems variationally as a functional of the
two-electron RDM, i.e., the PDF,  instead of the many-body
wavefunction. Both $\Psi$, and the 2-RDM are  unknown, but, unlike the wavefunction,
the 2-RDM  has the advantage that its application scales polynomially with the
number $N$ of electrons. However, the 2-electron RDM  must be a reduction of
an $N$-body wavefunction for it to be a physically acceptable 2-density.
Otherwise, the ground state energy for $N>2$ can even fall below the
 true ground state energy during a variational
calculation. So the 2-electron RDM  must be constrained to represent an
 $N$-electron  wavefunction.  Coleman
called these constraints $N$-representability conditions~\cite{Coleman63}.
The Hohenberg-Kohn minimization must be
 constrained to satisfy the requirements of
$N$-representability~\cite{Levy82, Mazziott12}. We do not present
the mathematical constraints enumerated by mathematicians here
as we will not use them. Instead, we propose to directly link our CHNC methods to
an underlying $N$-representable density.

The N-electron problem when treated in the grand canonical formalism uses a
chemical potential $\mu$ and an explicit number of electrons is used only
in the canonical ensemble. The passage to a canonical ensemble requires an
`inversion' of thermodynamic functions from the $\mu$ representation
 (see Ref.~\cite{PDW-XC84}). Such issues are irrelevant to the $N$-representability
problem as we can choose to work entirely in the canonical ensemble.

\subsection{Kinetic energy functionals}
The Hohenberg-Kohn method works directly with the density and does not
use a Kohn-Sham equation.
The implementation of DFT used in  Kohn-Sham theory~\cite{KS65}  maps
 the interacting electrons to a set of non-interacting electrons at
 the {\it interacting density},
 and calculates the `Kohn-Sham' one-electron wavefunctions $\phi_j(r)$
using some approximation, e.g.,  a local-density approximation (LDA)
 to the exchange-correlation potential based on the uniform electron fluid.
 Hence the corresponding
many-body wavefunction  is a single Slater determinant at $T=0$, and the Kohn-Sham theory
gives rise to the $N$-representable density $n(r)$ given by:
\begin{equation}
\label{density.eq}
n(r)=\sum_j|\phi_j(r)|^2f(\epsilon_j).
\end{equation}
At $T=0$ the Fermi occupation factors $f(\epsilon_j)$ are unity or zero for
occupied and  unoccupied states. Hence the summation at $T=0$ is over occupied states.

At finite $T$ the many body wavefunction $\Psi$ can be written as a sum over a set of Slater
determinants as done in the method of configuration interactions (CI). Thus, if there
 are $n$ electrons
  we need $N \gg n$ orthonormal functions,
e.g., Kohn Sham functions $\phi_j$ such that  the corresponding Fermi occupation
 factor of the highest energy state
used is deemed to be negligible. This is the statistical average over many
configurations, and the actual occupancies of the onebody states in each
electronic configuration are unity or zero. 
We can construct $N!/\{n!(N-n)!\}$ determinants out of the onebody states. 
 All these determinants contain (or omit) orbitals that  have unit (or zero) occupations.
 However, the squares of the  coefficients of these determinants, and the  occurrence
 of the orbitals in the determinants give rise to the fractional Fermi factors
contributing to the density in Eq.~\ref{density.eq}.

 Alternatively, the Fermi factors are simply given as the
statistical weights of the diagonal elements of the  $N$-representable 2-RDM constructed
 from $\Psi$ in the basis of $\phi_j$ one-body functions. 
 Hence, the finite temperature  case can also be re-stated as  a discussion in terms of
 properties of  Slater determinants,  as is the case for $T=0$.  In practical
 calculations at finite-$T$ the required basis sets become rapidly
  prohibitive as $T$ increases. 
Thus plane wave basis sets cut of at 500-1000 eV are needed, even with ultra-soft
pseudopotentials, in typical applications of DFT for
 warm-dense-matter~\cite{WittePOP18}.
CI calculations using $\Psi$ become impossible in such cases.

The classical map simplifies the DFT problem further and works with a classical electron
 system at a classical fluid  temperature $T_{cf}$. The latter is  constructed  to include the
 physical temperature $T$ as well as a kinetic-energy quantum correction brought in via a
 quantum temperature $T_q$ to be discussed below.

The KS $\phi_j(r), \epsilon_j$ have the physical meaning of being the eigenstates
 and eigenenergies of the fictitious non-interacting electron map of the interacting
 electron system, rather than those of  the original interacting electron system.
 The Kohn-Sham procedure guarantees the $N$-representability of the density by
 treating the kinetic energy operator explicitly,  without  using a kinetic energy (K.E.)
 functional as in Hohenberg-Kohn  DFT.

The simplest K.E. functional is used in Thomas-Fermi theory.  Extensions of
Thomas-Fermi theory under the name  of `orbital-free' DFT,  as well as practical applications
continue to be 
relevant~\cite{DG90,Carter00, KarasTrickyOF12,WhiteOF13, JCP-OrbitalFree2014, Clerouin15}.
 Many formulations use the von Weizs\"{a}cker ansatz where
just one orbital, viz., $\phi(r)=\surd{n(r)}$ is used in a Schr\"{o}dinger-like equation
 to obtain the kinetic energy. However, the
non-local nature of the K.E. operator continues to be a great stumbling block.
The excellent review by Carter~\cite{Carter00},  though littered with many acronyms,
shows the highly heuristic nature of the search for a K.E. functional that has
continued for some four score years.

Several exact requirements on the K.E. functional
(such as positivity) and their violation in various implementations
have  been noted~\cite{Levy88, Karasiev06}. However, whether `orbital-free'
 formulations lead to  $N$-representable densities, or  non-negative electron-electron
 pair-distribution functions etc., do not seem to have been studied.
In any case it is known that  calculations using  K. E. functionals
 are  far less accurate
than KS calculations. Furthermore, energies from such calculations
may fall {\it below} the exact energies, as the approximate K.E. functionals may not
satisfy  $N$-representability constraints. In fact, even some Kohn-Sham calculations that
use generalized  gradient approximations show such anomalies~\cite{Umrigar94}.

\subsection{The Neutral Pseudo Atom model}
A kinetic energy functional is unnecessary for simple `one-center' calculations which
are very rapid, and typical in atomic physics or with the neutral-pseudo-atom (NPA) model,
 originally proposed for
solids~\cite{Dagens75}, and adapted to finite-$T$ metallic fluids and  plasmas
~\cite{dwp82,ilciacco93,eos95,DWP-carb90,CPP-carb18}. The NPA has been
formulated in a number of different ways~\cite{ChiharaDRT92,AntaLouis00,XuHanson02,Chihara84}.
 Here
we follow the model of Ref.~\cite{eos95} which is a simplification of \cite{dwp82} and
adapted to multi-component finite-$T$ calculations. In these  NPA models of
 electron-ion systems,  the many-ion problem is replaced by a `one-ion' 
problem together with the corresponding ion-ion correlation functional, while the
 many-electron problem is replaced by a single-electron KS problem.

However, in simulations done with codes like the VASP~\cite{VASP} or ABINIT~\cite{ABINIT}
the many-ion problem is {\it not} reduced. Instead, they explicitly use some 100-200 nuclear
 centers, say $N_I$, and  even up to $N=N_e\sim 1000$ electrons in thousands of steps of
 KS and molecular-dynamics (MD) calculations. Hence such methods are extremely expensive and
 become prohibitive for many problems in warm-dense matter, materials science and
 biophysics. However, they provide useful benchmarks in simplified limits.
Such $N_I$-ion quantum calculations can be greatly simplified as follows.
\begin{enumerate}

\item  By the use of an explicit electron kinetic energy functional
  of the one-body electron density $n(\vec{r})$ if an adequate K.E.
  functional were available.

\item  Using a neutral pseudo-atom
 approach where the $N_I$ nuclei are replaced by a one-body ion density
 $\rho(\vec{r})$~\cite{dwp82, ilciacco93}, while the electrons are treated as usual
 as a functional of  $n(\vec{r})$ from KS theory.
 Since ions are normally classical particles,
  an ion is chosen as the origin of coordinates with no loss
 of generality.
Two coupled KS-equations for the  two subsystems ($e$-$i$) arise
on functional differentian  of $F$.

\begin{equation}
\label{npa2comp.eq}
\frac{\delta F([n],[\rho])}{\delta [n]}=\mu_e,\; \frac{\delta\ F([n],[\rho])}{\delta[\rho]}=\mu_I.
\end{equation}
The electron and ion chemical potentials appear on the RHS. The first equation reduces
to a one-center Kohn-Sham equation for the electrons in the field of the ion
at the origin, while the second equation defines  a classical distribution around the origin
containing an ion-correlation functional, and reduces to a
hypernetted chain (HNC) type integral equation~\cite{dwp82,ilciacco93}.
If there are many types of ions, a coupled set of one-center
HNC  equations appear~\cite{eos95}. 

This reduction of  the electron-ion problem  does not invoke the
 Born-Oppenheimer (BO) approximation, but BO can be
 implemented  by neglecting $F^{ei}_{xc}[n,\rho]$.
The solution of such one-center equations is numerically extremely rapid, even at finite $T$.
Such calculations reproduce the PDFs $g_{\rm cc}(r)$ of, say, molten carbon (or silicon)
containing a complex bonding structure that are only exposed by taking `snap shots'
in lengthy and expensive
DFT-MD simulations. That is, the {\it one-center} NPA calculations include sufficiently 
good ion-ion classical correlation functionals such that they are able, e.g.,  to
reproduce the peak in the $g_{\rm cc}(r)$ that corresponds to the 1.4-1.5\AA \ C-C
covalent bond as well the peaks in the $g(r)$ due to the hard sphere-like packing
effects seen in  DFT-MD simulations. This is demonstrated in
 Refs.~\cite{DWP-carb90, CPP-carb18}.

\item The NPA approach can also be further simplified by the use of a K.E. functional;
but the NPA calculation is  so rapid that little is gained on using approximate
K.E. functional with their own errors.
 \end{enumerate}

Several models use the  the name ``Neutral 
Pseudo Atom'',  but there are  significant differences. Thus Chihara uses
 a neutral-pseudo-atom construction where he begins from the HNC equation and
 identifies  a `quantum' Ornstein-Zernike (OZ) equation applicable to
 electrons as well~\cite{Chihara84}. 
Its  validity for quantum electrons is debatable. 
Thus Anta and Louis~\cite{AntaLouis00} in their implementation of an NPA
 using Chihara's `quantal HNC (qHNC)' scheme cautiously
avoid the  use an  {\it e-e} qHNC equation. The NPA approach proposed by
the present author and Perrot~\cite{dwp82} simply uses DFT for both
 electrons and ions, and
 invokes the HNC diagrams, bridge diagrams and the Ornstein-Zernike 
equation  only to construct an ion-ion correlation  functional \cite{dwp82, eos95}. 

A simplification of the effort to construct a kinetic energy functional is to look
for a classical description of the electrons. This is possible when the bound states
have already been treated using some complementary approach like the NPA, or
 when there are no
bound states because the system is highly compressed or at a temperature where
such effects can be neglected. If the system is a fluid or plasma, classical
integral equations or classical molecular dynamics can be used to obtain the PDFs
 of the classical-map  electrons that are not plagued by Fermion sign problems.

\subsection{The classical map hypernetted-chain scheme.}

The study of the electron distribution in a uniform electron liquid (UEL) when a
 `test electron' is placed at the origin  leads to the question of the
 direct calculation of the physically valid $g_{ee}(r)$ of the UEL rather than
 for a `test particle'. Here the electron kinetic energy functional must satisfy the
required constraints, and  also avoid any {\it selection} of an `electron' held at
 the origin whereby  it is made into a specific  `test particle'. Such a
problem does not arise for classical electrons~\cite{prl1}.  

We recapitulate the classical map hypernetted-chain (CHNC)
 scheme for the convenience of the reader. It has been used
 successfully~\cite{prl1,prl2,BulTan02,Totsuji09,LiuWuCHNC14}
 for a number of uniform systems, namely, 3D and 2D UELs, electron-proton
plasmas~\cite{hug02}, warm-dense matter~\cite{Bredow15}, 
 double quantum wells~\cite{lfc-dw19}) etc.
We present arguments to show that the pair-densities obtained via the
 classical-map technique  are $N$-representable.

Consider  an $N$ electron system in a volume $V$ such that $N/V=\bar{n}$,
forming a uniform  electron liquid in the presence of a neutralizing
positive uniform background. The electron eigenfunctions for the self-consistent field
 problem (Hartree as well as Hartree-Fock models) are simple plane waves.
\begin{equation}
 \phi_j(r)=\phi_{\vec{k}\sigma}(\vec{r})=(\bar{n}/N)^{1/2}\exp(i\vec{k}\cdot\vec{r})\zeta_\sigma.
\end{equation}
Here $j$ is an index  carrying any relevant quantum numbers including
 the  spin index $\sigma$ associated with the spin function $\zeta$,
with $\sigma=1,2$ or `up, down',  specifies the two possible spin states. 
Some of the vector notation will be suppressed for simplicity, as appropriate
for uniform liquids with spherical symmetry in 3D and planar symmetry in 2D.
The spin index may also be suppressed where convenient.
\subsection{The non-interacting pair-distribution function $g^0(r)$}
\label{g0.sec}
The many-electron wavefunction for non-interacting electrons in a uniform system,
 as well as for
 Hartree-Fock (mean-field)  electrons is a normalized antisymmetric
 product of planewaves~\cite{Mahantxt}, i.e, a Slater determinant $D(\phi_1,...\phi_j)$
of $N$-plane waves. Its square is the $N$-particle density matrix, while the
 PDF is the two particle density matrix~\cite{Vignale05}.
In the following we assume Hartree atomic units with $|e|=\hbar=m_e=1$, where
standard  symbols are used.
\begin{equation}
g_{\sigma_1.\sigma_2}(\vec{r}_1,\vec{r}_2)=V^2\Sigma_{\sigma_3\hdots\sigma_N}\int
d\vec{r}_3\hdots\vec{r}_N D(\phi_1,...\phi_j).
\end{equation}
If the spins are anti-parallel, then the non-interacting PDF, $g^0_{u,d}(r)$ is unity 
for all $\vec{r}$. Denoting $(\vec{r}_1-\vec{r}_2)$ by $\vec{r}$,
and $(\vec{k}_1-\vec{k}_2)$
by $\vec{k}$, we have, for  parallel spins,
\begin{eqnarray}
g^0_{\sigma,\sigma}(\vec{r})&=&\frac{2}{N^2}\Sigma_{\vec{k}_1,\vec{k}_2}f(k_1)f(k_2)
\left[1-\exp(i\vec{k}\cdot\vec{r})\right]\\
f(k)&=&\left[1+\exp\{(k^2/2-\mu^0)/T\}\right]^{-1}. 
\end{eqnarray}
Here we have generalized the result to finite $T$, where the temperature is measured in
energy units. Thus 
the non-interacting PDFs, i.e.,  $g^0(r)$ are explicitly available at $T=0$, and 
numerically at finite $T$.
\begin{eqnarray}
g^0_{\sigma,\sigma}(r)&=&1-F^2(r)\\
F(r)&=&(6\pi^2/k_F^3)\int f(k)\frac{sin(kr)}{r}\frac{kdk}{2\pi^2}\\
\mbox{3D, zero $T$}, \;   &=&3\frac{\sin(x)-x\cos(x)}{x^3},\;x=k_Fr. 
\end{eqnarray}
The equations contain the Fermi momentum $k_F$ which is defined in terms of
the mean density $\bar{n}$ and the corresponding electron Wigner-Seitz radius $r_s$.
Here we have assumed equal amounts of up and down spins (paramagnetic case) and
defined the Fermi wavevector  $k_F$.
\begin{equation}
k_F=1/(\alpha r_s),\;r_s=\left[3/(4\pi\bar{n})\right]^{1/3}, \;\alpha=(4/9\pi)^{1/3}.      
\end{equation}
 Similar expressions can be developed for
the 2D electron layer~\cite{prl2}, two coupled 2D-layers~\cite{lfc-dw19} 
or a two-valley 2D layer~\cite{2v2d04} relevant to silicon-metal oxide
 field effect transistors.  The method has also been used successfully to
obtain the local-field factors of 2D layers at zero and finite-$T$~\cite{LFC03}, and for
the study of thick 2D layers which are of technological interest~\cite{Thick2d05}.

\section{ The $N$-representability of  pair densities from the classical map}
We first discuss the non-interacting pair-density and then use its manifest
$N$-representability to establish the $N$-representability of the
interacting map.
\subsection{Non-interacting electron gas.}
\label{Nrep-g0.sec}
The PDFs $g^0_{\sigma,\sigma'}(r)=1-\delta_{\sigma,\sigma'}F(r)$ calculated in the previous
 section were derived from the Slater determinant 
$D(\phi_1\hdots,\phi_N)$ and hence manifestly $N$-representable. At this stage, irrespective
of where it came from,  we regard $g^0(r)$ as a classical pair-distribution function for
classical electrons interacting by a classical pair potential 
$\beta\mathcal{P}(r)$ where $\beta$ is the inverse temperature . This is the first step in
 our  classical map, and we may now identify one of the classical particles as being at
 the origin, without loss of generality, in a classical picture of the PDF. Clearly, for 
anti-parallel spins, i.e., $\sigma\ne\sigma'$, the pair-potential $\beta\mathcal{P}(r)$ is
 zero, while it is finite and creates the well-known `exclusion hole' in the PDF of
 two parallel-spin particles. Hence $\mathcal{P}(r)$ has been called the `Pauli exclusion
 potential' and should not be confused with the Pauli kinetic potential that appears in
 the theory of the kinetic energy functional.

 F. Lado was the first to present an  extraction of $\beta\mathcal{P}(r)$ for 3D
 electrons at $T=0$ using  the hypernetted-chain (HNC) equation and the
 Ornstein-Zernike (OZ)  equation~\cite{Lado67}.
 Only the dimensionless potential, $\beta\mathcal{P}(r)$ is determined
 from the equations. Although the physical temperature $T$ of the quantum fluid
 is zero, the temperature of the classical fluid  invoked by the map is left
 undetermined (but nonzero) in the `non-interacting' system. The Pauli
exclusion potential for 2D electrons at arbitrary $T$ was derived in Ref.~\cite{prl2}.
Although the quantum electrons are not interacting via a Coulomb potential, 
 $\beta\mathcal{P}(r)$ becomes a classical  manifestation of entanglement
 interactions  which scale as $r/r_s$, and hence extend  to arbitrarily large
 distances~\cite{cdwEntang13}. Assuming that $g^0(r)$  can be written as an HNC
 equation, we have:
\begin{eqnarray}
g^0(r)&=&\exp\left[-\beta\mathcal{P}{r}+h^0(r)-c^0(r)\right]\\
h^0(r)&=&c^0(r)+ \bar{n}\int d\vec{r}\,'h^0(|\vec{r}-\vec{r}\,'|)c^0(\vec{r}\,')\\
h^0(r)&=&g^0(r)-1.
\end{eqnarray}
The first of these is the HNC equation, while the second equation is the Ornstein-Zernike
 relation. These contain the direct correlation function $c^0(r)$ and the total correlation 
function $h^0(r)$. It should be noted that we have ignored the two-component character
of the electron fluid (two spin types) in the equations for simplicity, but the full
expressions are given in, say, Ref.~\cite{prl1}. These equations can be solved by taking
their Fourier transforms, and the Pauli exclusion potential can be obtained by the
inversion of the HNC equation. 
The ``Pauli exclusion potential'' (PEP)  $\beta{\cal{P}}(r)$ is given by
\begin{equation}
\label{paudef.eq}
\beta {\cal{P}}(r) = -log[g^0(r)]+h^0(r)-c^0(r).
\end{equation}
The PEP is a
universal function of $rk_F$ or $r/r_s$. It is long ranged and
mimics the exclusion effects of Fermi statistics. At finite $T$
its range is about a thermal de Broglie wavelength and is increasingly
hard-sphere-like as $r\to 0$ . The Fourier transform $\beta\mathcal{P}(q)$
in 3D  behaves as $\sim 1/q$ for small $q$, and as $\sim
 c_1/q^2+c_2/q^4$ for large $q$.

\begin{figure}[t]
\includegraphics[scale=0.3, trim=0.5cm 0.5cm 0.5cm 0.5cm, clip=true]{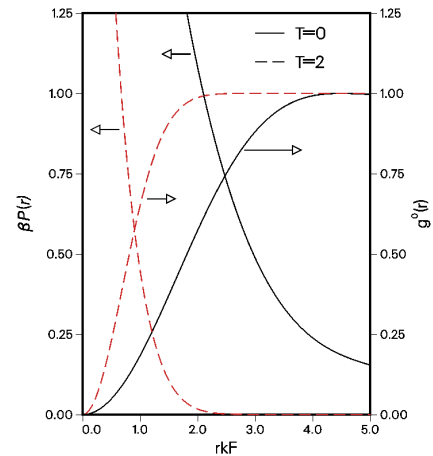}
\caption{The exclusion potential, Eq.~\ref{paudef.eq}, and the noninteracting
PDF, $g^0_{\sigma,\sigma}$ at $t=T/E_F=0$ (solid line) and at $t=2$ (dashed line).
They are universal functions of $r/r_s$. The PDF $g^0_{\sigma\ne\sigma'}(r)=1$
 as there is no exclusion effect for $\sigma\ne\sigma'$.} 
\label{Pauli.fig}
\end{figure}

Plots of $\beta \mathcal{P}(r)$ and $g^0(r)$ for a 3D UEL are
given in Fig.~\ref{Pauli.fig}. 

We note that the  HNC or MHNC integral equation, together with the OZ equation
may be regarded as a transformation where, given the dimensionless pair potential
 $\beta \phi_{ij}(r)$, the corresponding PDF, i.e.,  $g_{ij}(r)$ is generated.
 Similarly, given the $g_{ij}(r)$,
HNC inversion is the process which extracts the corresponding $\beta \phi_{ij}(r)$.
The value of $g(r)$ for {\it the full range} of $r$, or additional constraints are needed
to obtain an unequivocal HNC inversion to extract a valid pair
 potential from a PDF~\cite{CDW-LWR86,Rosen-Kahl97}.

\section{The interacting  system and its classical map}
\label{interacting.sec}
In the previous section we reviewed  a classical fluid whose $g^0(r)$ exactly recovers the
PDFs of the non-interacting quantum UEL at any density, spin polarization  or temperature.
From now on, for simplicity we consider a paramagnetic electron liquid (equal amounts of up
spins and down spins) although spin-dependent quantities will be indicated where needed
 for clarity. Although the quantum liquid was `noninteracting', the classical map already
 contains the pair potential $\beta U_{ij}$= $\beta\mathcal{P}(r)$.

On addition of a Coulomb interaction $\beta V_{ij}(r)$ the total pair potential becomes
\begin{equation}
\label{pairpot.eq}
\beta U_{ij}(r)=\beta\mathcal{P}(r)+\beta V_{cou}(r).
\end{equation}
The temperature $T=1/\beta$ occurring in Eq.~\ref{pairpot.eq} is as yet unspecified.
In quantum systems the Coulomb interaction is given by the {\it operator}
 $1/|\vec{r}_1-\vec{r}_2|$
which acts on the eigenstates of the interacting pair. It can be shown
 (e.g., by solving the relevant quantum scattering equation) that the classical
Coulomb interaction, $V_{cou}(r), r=|\vec{r}_1-\vec{r}{_2}|$  acquires a diffraction
 correction  for close approach. Depending on the temperature $T$, an electron is
localized to within a thermal de Broglie wavelength. Thus, following earlier work
on diffraction corrected potentials, (e.g., in Compton scattering in
high-energy physics), or in plasma physics as in, e.g., 
 Minoo {\it et al.} \cite{minoo81}, we use a ``diffraction
corrected''  potential.
\begin{equation}
\label{potd}
V_{cou}(r)=(1/r)[1-e^{-r/\lambda_{th}}];\;
\lambda_{th}=(2\pi\overline{m}T_{cf})^{-1/2}.
\end{equation}
Here $\overline{m}$ is the reduced mass of the interacting electron pair, i,e,
 $m^*(r_s)/2$ a.u., where $m^*(r_s)$ is the electron effective mass.  It
is weakly $r_s$ dependent, e.g, $\sim$0.96 for $r_s$ = 1. In this
work we take $m^*$=1. The ``diffraction correction'' ensures the
correct quantum behaviour of the interacting $g_{12}(r\to 0)$ for all $r_s$.
The essential features of the classical map are
\begin{enumerate}
\item The use of the exact non-interacting quantum PDFs $g^0_{\sigma,\sigma'}(r)$ as inputs.
\item A diffraction corrected Coulomb interaction. 
\item The specification of the temperature of the
classical Coulomb fluid $T_{cf}(r_s)=1/\beta$ as the one that recovers
 the quantum correlation energy $E_c(r_s)$.
\end{enumerate}
The selection of $T_{cf}$ is a crucial step. This is guided
by the Hohenberg-Kohn-Mermin property that the exact minimum free energy is
determined by the true one-body electron density $n(r)$. Since we are dealing
with a uniform system,  the Hartee energy $E_{\rm H}$ is
zero. The exchange energy $E_x$ is already correctly accounted for
by the construction of the classical-map $g^0(r)$ to be identical with the quantum $g^0(r)$
at any $T$ or spin polarization. Hence the only energy to account for is $E_c$. So we choose
 to select the temperature $T_{cf}$ of the
classical Coulomb fluid to recover  the known DFT correlation energy $E_c$ at each
 $r_s$ at $T=0$. Since this is most accurately known for the spin-polarized electron
 liquid, $T_{cf}$ is best determined from $E_c(r_s)$ for full spin polarization.
A trial temperature is selected and the interacting $g(r,\lambda)$ is determined
for various values of the coupling constant $\lambda$ in the interaction $\lambda V_c(r)$
 to calculate a  trial $E_c$ at the given $r_s$ from the coupling constant integration.
 The temperature  is adjusted until the $E_c(r_s,T_{cf})$ obtained from the
 classical fluid $g(r)$  agrees with the known quantum $E_c(r_s,T=0)$. Given an
 electron fluid at $T=0$, the temperature of the classical fluid with the same
 $E_c$ is called its {\it quantum temperature} $T_q$. This was parametrized as:
\begin{equation}
T_q/E_F=1.0/(a+b{\surd{r_s}}+cr_s)
\label{tfit}
\end{equation}
For the range $r_s=1$ to 10, $T_q/E_F$ goes from 0.768 to 1.198. The values
of the parameters $a,b,c$ are given in Ref.~\cite{prl1}.

There is no {\it a priori} reason that the $n(r)$, i.e., $\bar{n}g(r)$
obtained by this procedure would agree with the quantum $\bar{n} g(r)$,  except for the
Hohenberg-Kohn theorem that requires $n(r)$ to be the true density distribution when the
energy inclusive of the XC-energy is correctly recovered. Many well-known
and often  very useful  quantum procedures
 (e.g., that of Singwi et  al.~\cite{STLS68, VS72}) for the PDFs lead to negative
 $g(r)$ as $r_s$ is  increased beyond unity even into the `liquid metal' $r_s$ range.  

However, as shown in Refs.~\cite{prl1,prl2,PDWXC} etc., the classical map
 HNC $g(r)$ is an accurate approximation to the QMC PDFs then available only at $T=0$. 
Correlations are stronger in reduced dimensions. The classical map for the 2D UEL
was constructed using the modified-HNC (MHNC) equation where a hard-sphere bridge
 function
 was used, with the hard-sphere radius determined by the Gibbs-Bogoliubov criterion,
as given by Lado, Foils and Ashrcoft (LFA)~\cite{LFA83}.  Other
 workers~\cite{BulTan02, Totsuji09,LiuWuCHNC14}  have examined different parametrizations
 than our  Eq.~\ref{tfit}. Datta  and Dufty~\cite{SandipDufty13} examined
 the classical map approach  and the method of quantum statistical
 potentials~\cite{Kelbg63,Filinov04} within a grand-canonical formalism.
 They  proposed using additional conditions  (besides the requirement that $E_c$ is
 reproduced by $T_{cf}$) to constrain the classical  map for warm dense electrons,
 a topic recently reviewed by Dornheim {\it et al.}~\cite{Dornheim18}.

Although $E_c$ values at $T=0$ were available when the classical map for the UEL was constructed,
 no reliable XC-functional (beyond RPA) was available for the finite-$T$ electron liquid. 
Hence we proposed the use of the `$T$ ansatz':
\begin{equation}
\label{fT.eq} 
T_{cf}=(T_q^2+T^2)^{1/2}
\end{equation}
 as a suitable map for the finite-$T$ UEL. This was based on the behaviour
 of the heat capacity and  other thermodynamic properties of the UEL. Furthermore,
using Eq.~\ref{fT.eq} it became possible to predict the XC-free energy
 $F_{xc}(r_s,T)$ as well as the finite-$T$ PDFs of the UEL at arbitrary temperatures
 and spin polarizations.  These were found to  agree closely with the $F_{xc}(r_s,T)$
and PDFs resulting from the Restricted  Path Integral Monte Carlo (RPIMC) simulations
reported 13 years later by Brown {\it et al}.~\cite{BrownXCT13}.
 The Brown {\it et al.} data have been
used by Liu and Wu~\cite{LiuWuCHNC14} to construct a direct fit of a $T_{cf}$ that
 avoids the model used  in Eq.~\ref{fT.eq}, by using temperature dependent parameters
 $a, b$ and $c$ in Eq.~\ref{tfit}. The RPIMC data have been parametrized by Karasiev
 {\it et al.},  Ref.~\cite{KSDT14}. However, 
Groth, Dornheim et  al.~\cite{GrothFxc17} presented a
 new {\it ab initio}  parametrization of  $F_{xc}(r_s,T)$  using accurate
 data  from recently developed  finite-$T$  fermionic PIMC  methods that deal
with the sign problem more carefully and also
compensate more systematically for finite-size effects~\cite{Dornheim18}.
 These agree even more closely
with the  CHNC data.

\begin{figure}[t]
\centering
\includegraphics[width=7cm]{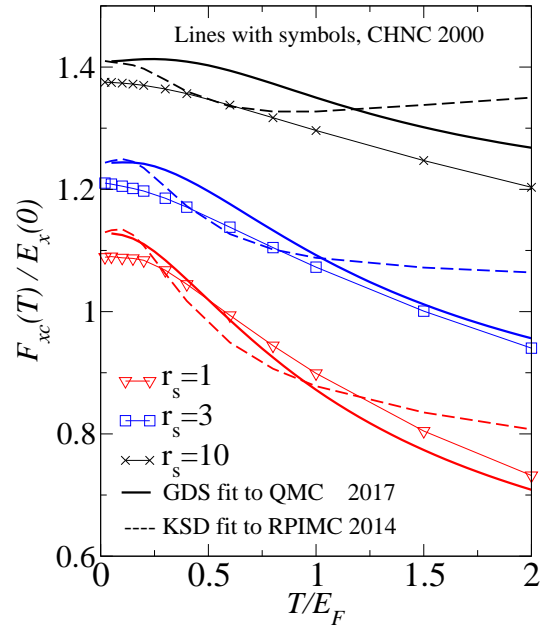}
\caption{(Color online)
Finite-$T$ exchange and correlation free energy $f_{xc}(r_s,T)$ 
scaled by the exchange energy $E_X=F_X$ at $T$=0 as a function of  the
reduced temperature $t=T/E_F$ in units of the Fermi energy is displayed. The lines with
 symbols  are results from CHNC calculations~\cite{PDWXC}.
 The restricted path integral Monte Carlo (RPIMC) data of Brown {\it et al}.,
 Ref.~\cite{BrownXCT13},
 as parametrized by Karasiev {\it et al.}~\cite{KSDT14} are shown as dashed lines.
 The thick continuous lines are from the Groth {\it et al}. parametrization of very
accurate fermionic-PIMC  data. The temperature range $t <1$ is
relevant to WDM studies. 
}
\label{fxc.fig}
\end{figure}

Calculations of $F_{xc}$ using the finite $T$ parametrization given by
Perrot and Dharma-wardana~\cite{PDWXC} are compared with the Karasiev {\it et al.} 
parametrized results, and those of Groth {\it et al.} in Fig.~\ref{fxc.fig}. The
 classical temperature ansatz of Eq.~\ref{fT.eq} recovers the
highly accurate Groth {\it et al}. results to within 94 \%, i.e., with an error of
at most 6\%. The parametrizations 
given by PDW~\cite{PDWXC}, Iyatomi and Ichimaru, and
 subsequent parametrizations incorporate the high-$T$  Debye-H\"{u}ckel
 limit of $F_c$, the high-$T$  behavior  of $F_x(T)$, as well  as the behaviour at
 the $T=0$ limit. The PDW fit to the CHNC data fall below the Groth
{\it et al.} data near $T=0$ partly because  older $T=0$ data were used in
the CHNC parametrizations. The CHNC method has also been used to
construct $F_{xc}(T)$ for 2D electron layers and used to
calculate finite-$T$ local-field factors, PDFs, and related quantities
relevant to double quantum wells, metal-oxide field effect transistors and
nanostructures; but no finite-$T$ QMC or PIMC bench marks are currently available for
the 2D electron system.   

\subsection{$N$-representability of the interacting $g(r)$ of the classical map.}
\label{Nrep-g.sec}
The conditions $n(r)=\bar{n}g(r)>0$, and $\int n(\vec{r})d\vec{r}=N$ are
always satisfied by  the classical map. Furthermore, the classical map becomes
more accurate as $t=T/E_F$ is increased, or when $r_s$ is increased, since
quantum electrons become increasingly classical in those limits.

We present two types of arguments to conclude that the $g(r)$ of the interacting
UEL obtained by the classical map is $N$-representable. One of them is
a formal argument based on CHNC being a``well behaved'' transformation of the already
$N$-representable non-interacting  density. The second is a
 practical demonstration of the implementability of CHNC method and the
close agreement with results from
 QMC, PIMC and  other more microscopic  benchmark
 calculations. Finally, we give an example of a CHNC calculation for electron-hole
 layers at finite temperatures, as an example of topical technological interest
 for which  QMC, PIMC, and even DFT seem to be quite
 prohibitive at present.

(1){\bf Argument based on the HNC equation  being an
 $N$-representability  conserving transformation.}\\
Once the $g^0(r)$ of the quantum fluid is evaluated we consider a classical fluid
which has the same $g^0(r)$.  
The non-interacting $g^0(r)$ and the corresponding $n^0(r)=\bar{n}g^0(r)$
 of the classical fluid are  generated from the homogeneous density $\bar{n}$ by a
 transformation where the origin
of coordinates is moved to one of the particles. The corresponding transformation
of the density profile  is
 written as:
\begin{eqnarray}
n^0(r)&=&\mathcal{T}_0(r)\bar{n}\\
\mathcal{T}_0(r)&=&\exp\left[\beta\mathcal{P}(r)+h^0(r)+c^0(r)\}\right].\\
\end{eqnarray}
The so generated $n^0(r)$ is $N$-representable by its construction from a Slater determinant.
Then, in a next step the  interacting $g(r)$ is generated from the $N$-representable
non-interacting  $g^0(r)$ by a transformation
 which can be written as:
\begin{eqnarray}
g(r)&=&\mathcal{T}_1(r)g^0(r)\\
\label{T1.eq}
\mathcal{T}_1(r)&=&e^{\left[\beta V_{cou}(r)+\{h(r)-h^0(r)\}+\{c(r)-c^0(r)\}\right]}\nonumber\\
           &=&\exp\left[\beta \{V_{cou}(r)+V_{\rm MF}(r)+V_{\rm xc}(r)\}\right].
\end{eqnarray}
In effect, the uniform density $\bar{n}$ has been transformed (by a selection of the origin
of coordinates, and by switching on the Coulomb interaction) by a single composite
 transformation
$\mathcal{T}=\mathcal{T}_1\mathcal{T}_0$ 
with its components acting one after the other.

In  equation \ref{T1.eq} we use  $V_{\rm xc}(r)$ to indicate the exchange-correlation
 correction  to the mean-field potential $V_{\rm MF}(r)$ as discussed in
 Ref.~\cite{dwp82} where explicit expressions for these classical XC-potentials
in the HNC approximation are given. These potentials are  expected to be
 well-behaved functions. 
The diffraction-corrected  classical Coulomb  potential $V_{cou}(r)$  has a finite-value
 at $r=0$, and not singular,  unlike the point-Coulomb  potential $1/r$ which is not used
in CHNC. Hence we may regard the above transformation as  being mathematically equivalent
 to a type of smooth, or `well-behaved' coordinate transformation of $\vec{r}$ to
  another variable $\vec{s}$.  
\begin{equation}
d\vec{s}=\mathcal{T}(r)\bar{n}d\vec{r} =n(r)d\vec{r}.
\end{equation}
That is, the initial planewave states $(\bar{n}/N)^{1/2}\exp(i\vec{k}\cdot\vec{r})$ are
transformed to a new set $(n(\vec{r})/N)^{1/2}\exp(i\vec{q}\cdot\vec{s}(r))$. It is easily
shown that they form a  mutually orthogonal complete set. For instance, consider the initial 
planewave state  used in the Slater determinant, i.e., $\phi_j(\vec{r})=\phi_k(\vec{r})$ 
and consider its transformed state $\tilde{\phi}_k(\vec{r})$ given below: 
\begin{eqnarray}
\phi_k(\vec{r}) &=& (\bar{n}/N)^{1/2}\exp(i\vec{k}\cdot\vec{r})\\
\tilde{\phi}_k(\vec{r})&=& (n(\vec{r})/N)^{1/2}\exp(i\vec{k}.\vec{s}(\vec{r})).
\end{eqnarray}
We regard $\vec{k}$ as an arbitrary $k$-vector and hence it is sufficient to
 transform $\vec{r}$, while
the theory can also be constructed entirely in $k$-space in an analogous manner.
 The transformed  wavefunctions $\tilde{\phi}_k(\vec{r})$ have the following properties:
\begin{eqnarray}
\int \tilde{\phi}^*_{k'}(\vec{r}) \tilde{\phi}_k(\vec{r})d\vec{r}&=&\int\frac{n(\vec{r})}{N}
 e^{i(\vec{k}'-\vec{k})}d\vec{r}\\
     &=&\frac{1}{N}\int \exp\{i(\vec{k}'-\vec{k})\} \frac{d\vec{s}}{N}\\
     &=&\frac{(2\pi)^3}{N}\delta^3(\vec{k}'-\vec{k}).
\end{eqnarray}
Furthermore,
\begin{equation}
\int \tilde{\phi}^*_k(\vec{r})\tilde{\phi}_k(\vec{r}\,')\frac{d\vec{k}}{(2\pi)^3}
=\frac{\delta^3(\vec{r}-\vec{r}\,')}{N}.
\end{equation}
Hence the transformed functions $\tilde{\phi}_k(\vec{r})$ form a complete orthogonal set.
 This implies that the initial Slater determinant $D(\phi_{k_1},\hdots,\phi_{k_N})$ of
 the  noninteracting electron system transforms
 to the determinant $D(\tilde{\phi}_{k_1},\hdots,\tilde{\phi}_{k_N})$ of the
 interacting system, explicitly showing the
 $N$-representability of the $n(r)=\bar{n}g(r)$ obtained via the classical map
 which consists  of the application of the two transformations
 $\mathcal{T}_1\mathcal{T}_0$. Furthermore, the
 transformations commute, in the sense that one may first apply {\it only} the
 diffraction
corrected Coulomb potential to non-interacting fermions to generate a $g^c(r)$ for
 a Coulomb fluid, and then apply the Pauli  exclusion potential to generate the
 fully interacting classical  map inclusive of exchange-correlation effects,
 or {\it vice versa}. This is equivalent to iterating the
HNC equations from the non-interacting state via two different paths, and indeed the two
 different procedures, $\mathcal{T}_1\mathcal{T}_0$ and $\mathcal{T}_0\mathcal{T}_1$
 lead to  the same final $g(r)$. 

In the above demonstration we have appealed to the concept of ``well behavedness'' of the
potentials. These have been defined by Lieb to be based on a Hilbert space of potentials
$\mathcal{V}=L^3/2+L^\infty$~\cite{Lieb83}. These are potentials that do not become
 singular or include infinite barriers, discontinuities  etc. The potentials used in
CHNC  satisfy  such concepts of ``well behavedness''.

What if a phase transition, e.g., a Wigner crystallization,
intervenes in passing from the non-interacting to the interacting system?
 The CHNC procedure
for a uniform system will smoothly proceed to the best interacting fluid
 state (an upper bound to the ground-state energy)  and not the lower-energy Wigner
 crystal state. This is still consistent with the variational principle and
 $N$-representability. 

The above analysis confirms the $N$-representability of the pair-densities of the
interacting uniform electron liquid  generated by the classical map presented here.
A similar analysis can be given if MD were used to generate the interacting $g(r)$ via
CHNC potentials, instead of using the HNC equation.

(2){\bf Argument based on the $N$-representability of the QMC density.}\\
The diffusion quantum Monte Carlo (DQMC)  calculations use a Slater determinant together
with Jastrow factors, and hence the DQMC procedure is based on an explicit
 many-body wavefunction  whose variation produces a minimum energy and a
 corresponding $E_c(r_s)$. Hence
its two-particle reduced density matrix, i.e., the electron-electron PDF is
 $N$-representable;  the correlation energy $E_c$  associated with the $N$-representable
 two-body density is the  correlation contribution to the best approximation to the energy
 minimum as per Hohenberg-Kohn theorem,  since the minimum is
 achieved only for the true density.  The CHNC electron-electron PDFs agrees with the
 DQMC-$g(r)$  with {\it no attempt at fitting} the PDFs. The only input is  the single
 number $E_c(r_s)$ at each density introduced via the classical temperature $T_{cf}$ - a
 classical kinetic energy. The CHNC electron density $\bar{n}g(r)$ agrees closely
at every $r$ with those of the $N$-representable density $\bar{n}g_{\rm qmc}(r)$. Further
more, the CHNC energies $E_{xc}(r_s,\zeta$) at arbitrary spin polarizations $\zeta$
that were not include in any fits agree with microscopic calculations. At finite-$T$,
 an ansatz is used for $T_{cf}$ and yet the agreement is good to within 94\%. Hence we
 conclude that the  CHNC $n(r)$ is  as $N$-representable  as the DQMC procedure.

The classical pair potential $U(r)=\mathcal{P}(r)+V_{cou}(r)$ may
 be used in a classical molecular dynamics simulation to generate the interacting $g(r)$
 instead of using the HNC equations. Such a procedure can reveal crystal ground states
and go beyond the liquid-model inherent in the usual HNC equations.
 It is also possible to generate the dynamics of fluid states, e.g., determine
$S(k,\omega)$ by a classical simulation. However, the Pauli exclusion interaction is
 really a kinematic quantum effect and not a true `interaction'. It is not known
at present  whether such a classical map $S(k,\omega)$ agrees in detail with the
 quantum $S(k,\omega)$ for an  interacting electron fluid. 
 
\section{CHNC method for systems of interacting electrons and ions}
\label{two-comp.sec}
In this section the UEL model is extended to two interacting subsystems, namely electrons and
ions, or possibly for electrons and holes. The application of the CHNC method to  coupled 
electron-ion system will be
 illustrated by calculations of hydrogen plasmas where the CHNC results are compared
 with NPA calculations, as well as
recent QMC, PIMC, DFT-MD and other $N$-center simulation methods. The CHNC, NPA and
 Quantum-simulation methods are in good agreement. This agreement is the basis of our
second argument for the $N$-representability of the pair-densities obtained from the
CHNC method.
  
Consider the two coupled density functional equations of the NPA, viz.,
Eqs. \ref{npa2comp.eq}. As shown in the appendix to Ref.~\cite{dwp82} these equations for
$n(r),\rho(r)$, when applied to classical particles reduce to classical Kohn-Sham equations
which are Boltzmann like distributions. In the classical case they can be
reduced to two coupled HNC-like equations for the electrons and ions. The HNC equation
(with or without a Bridge term) for the electrons when replaced by their classical map
 gives  the CHNC  equation which now includes  an electron-ion contribution to the
 potential of mean force.  Similarly, the HNC-like equation for the ions will contain
 contributions from the electrons. That is,  the electron screening of the ions, or ion
 screening of the electrons is controlled by the CHNC equations,
 which  only needs the basic pair interactions. If the particles are in the quantum
 regime, a Pauli-exclusion potential is needed, be it for electrons, or for protons (or holes
in semiconductor applications). 

As an example, we take a system of ions of mean charge $\bar{Z}$, mass $M$, with a mean
 density $\bar{\rho}$ interacting with a  system of electrons at a
 mean density $\bar{n}=\bar{Z}\bar{\rho}$. The electron mass is unity (atomic units).
For simplicity we assume that there is just one kind of ion, and that
 $\bar{Z}=1$ as for a hydrogenic
 system. We denote the ion species  by p$=H^+$. 
The coupled CHNC equations are given in Eq.~\ref{hnc} and in
Refs.~\cite{hug02,Bredow15}.  They are  discussed below.

The densities $\bar{\rho}$ and $\bar{n} $ are equal since
 the ion charge $\bar{Z}$ = 1.
Consider a fluid of total  density $n_{tot}$, with
three species, electrons of two types of spin, $n_s=\bar{n}/2$ for
$s=1,2$ for the two spin species, and $s=3, n_3=\bar{\rho}$ for the ions,
denoted here as `p'. 
The physical temperature is  $T$, while the classical-fluid temperature
of the electrons,
$T_{cf}=T_{ee}= 1/\beta_{ee}$, with
with  $1/\beta_{ee}$= ${\surd{(T^2+T_q^2)}}$.
For  the ion p=$H^+$, no quantum correction is usually needed and
 $T_{\rm pp}=1/\beta_{\rm pp}$ is  $T$. Otherwise an ion-quantum
temperature $T_{q\rm p}$ is defined as before, using Eq.~\ref{tfit},
but using the Fermi Energy $E_{F\rm p}$ of the ions (or holes). If the densities are those
typical of stellar densities, then the calculation will automatically
be for quantum hydrogen ions as appropriate. Normally, treating the positively
charged counter particles of the electrons quantum mechanically is
not needed except in semi-conductor structures.

Thus the classical map converts the system with the physical temperature $T$
into a system with two temperatures $T_{ee},T_{\rm pp}$ associated with the
 two different subsystems. It should be noted that the temperature is not an
observable in pure quantum systems as there is no operator associated with it.
But  it  a  Lagrange multiplier  in quantum {\it statistical} systems
 and ensures the conservation of the energy of each subsystem when coupled to
classical heat baths where it is measurable.
Thus two Lagrange multipliers are  implied in the electron-ion system.
However, the cross subsystem temperature $T_{e\rm p}$  is not easily defined
as there is no uniform density or specific  energy associated with the
 cross-interactions, and no conserved quantity to define a Lagrange multiplier.
 It can however be unambiguously extracted from an
 NPA calculation or constructed from a suitable physical model, as discussed below.
 So, using $T_{ij}=1/\beta_{ij}$ and $\phi_{ij}$
for the interparticle temperatures and pair-potentials, the coupled
CHNC and OZ  equations for the PDFs are:
\begin{eqnarray}
\label{hnc}
g_{ij}(r)&=&e^{[-\beta_{ij}\phi_{ij}(r)
+h_{ij}(r)-c_{ij}(r) + B_{ij}(r)]}\\
 h_{ij}(r) &=& c_{ij}(r)+
\Sigma_s n_s\int d{\bf r}'h_{i,s}
(|{\bf r}-{\bf r}'|)c_{s,j}({\bf r}')
\label{oz}
\end{eqnarray}
The pair potential $\phi_{ij}(r)$ between electrons  is
just the diffraction corrected Coulomb potential  $V_{cou}(r)$
added to the Pauli exclusion potential which vanishes as the
de Broglie radius of the particles becomes negligible in approaching
the classical regime.  The interaction between two ions is also a Coulomb
potential with a diffraction correction with a very small length scale
($\sim 1/\sqrt{M}$) due to the mass $M$ of the ions.

The HNC is sufficient for the uniform 3-D electron liquid  for a
range of $r_s$, up to $r_s=50$, as shown previously~\cite{PDWXC}.
But the bridge term is important for the ions at high compressions
and low temperatures. Hence we neglect the {\it e-e} bridge corrections  in
this study, but retain them for ions.

The construction of $\beta_{e\rm p}, \phi_{e\rm p}(r)$ requires attention.
Only the product, $\beta_{e\rm p}\phi_{ij}(r)$ is unambiguously available
from the theory, and that is necessary and sufficient for CHNC calculations.
However, simple electron-ion interaction models may also be used.
Thus  Bredow {\it et a.l}~\cite{Bredow15}
examined the applicability of a very simple electron-ion
interaction and a simple model for the inter-subsystem
temperature of the two-temperature classical map. 
\begin{eqnarray}
\label{Bredow.eq}
\phi^0_{e\rm p}(r)&=&U(r)\{1-\exp(-r/\lambda_{e\rm p})\},\\
U(r)&=&(-\bar{Z}/r)f(r),\; f(r)=1. \\
\lambda_{e\rm p}&=&(2\pi\overline{m}T_{e\rm p})^{-1/2},\; \overline{m}=1.,\\
T_{e\rm p}&=&(T_{ee}T_{\rm pp})^{1/2}.\label{Tep.eq}
\end{eqnarray}
The e-p de Broglie length $\lambda_{ep}$ moderates
the $r\to 0$ behaviour of the electron-proton interaction. The latter
includes a pseudopotential $U(r)$ where the form factor $f(r)$ is set to
one for the liner-response regime. For ions with a bound core of radius $r_c$, the form
factor can be chosen to have the Heine-Abarankov form. In such cases the diffraction
correction becomes irrelevant. 

Equation~\ref{Tep.eq} for $T_{e\rm p}$ as a geometric mean of the
inter-subsystem temperatures is justifiable in the large-$r$ or small-$k$ limit
 using compressibility sumrule arguments~\cite{Shaffer17,Bredow13}.
 However, for small-$r$,
 binary collisions  dominate and the ansatz becomes less valid.
 Furthermore the electron density
near the nucleus is large, and the effective $T_{ee}$ increase from the bulk
value. Hence a single $T_{e\rm p}$ is not strictly possible but found to
work quite well, as shown below. The ion-ion PDF is insensitive to the choice
of $T_{e\rm p}$, and hence the model, Eq.~(35), proves to be very convenient.

\begin{figure}[t]
\centering
\includegraphics[width=8cm]{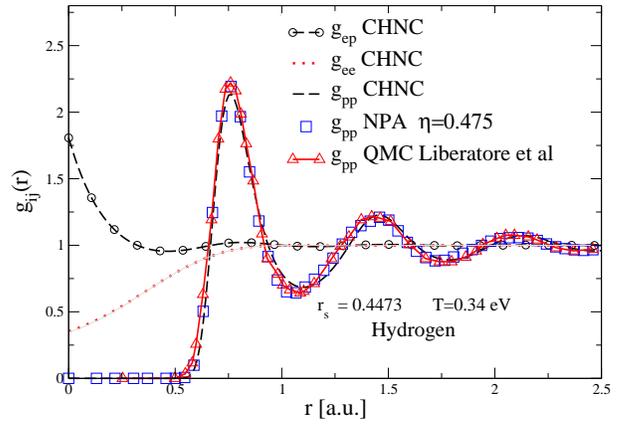}
\caption {Pair distribution functions $g_{ij}(r),i=e,{\rm p}$ using the
simple CHNC model $T_{e\rm p},\phi_{ep}$ from Eq.~\ref{Bredow.eq}, from the NPA,
 and from  QMC (Liberatore {\it et al.}~\cite{Liberatore11}), for
fully ionized hydrogen. The ion density is  $\bar{\rho}$ =
 1.8$\times10^{25}$ cm$^{-3}$ (i.e,$\sim$ 350 times the density of solid hydrogen)
with $T=0.34$ eV. The NPA, CHNC and QMC agree very well for
$g_{\rm pp}$ when Bridge corrections specified by the hard-sphere packing
fraction $\eta$ are
included in the NPA. See Fig.~\ref{nei.fig} for more details on $g_{e\rm p}$
}
\label{BredowH.fig}
\end{figure}

\begin{figure}[t]
\centering
\includegraphics[width=8cm]{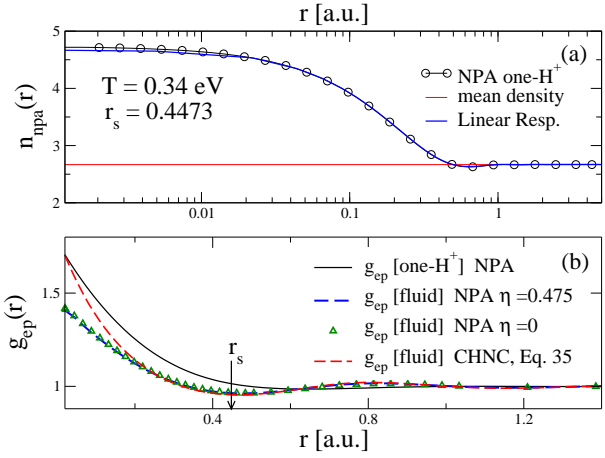}
\caption {(a)The free-electron density $ n_{\rm npa}(r)$ calculated from the
NPA model at  a proton in a hydrogen plasma, $T$ = 0.34 eV, with an ion density of
 $\bar{\rho}$ = 1.8$\times10^{25}$ cm$^{-3}$, i.e., $r_s=0.4473$. At this high
density linear-response theory (dashed curve) is accurate. (b)The  density
displacement can be used to define $g_{e\rm p}(r)$ for the one-proton system
and its generalization to the fluid. Results for $g_{e\rm p}$ obtained from a
 simple model of Eq.~\ref{Bredow.eq} within the CHNC, and calculations
from the NPA for hydrogen fluid using the MHNC equation are displayed.
}
\label{nei.fig}
\end{figure}

We display the results from the simple model, Eq.~\ref{Bredow.eq} in
Fig.~\ref{BredowH.fig} and compare them with the results from NPA calculations
as well as with highly computer-intensive QMC  simulations by
 Liberatore {\it et al.}~\cite{Liberatore11}. Liberatore {\it et al.}
assume a linear-response form for the proton-electron interaction
following the Hammerberg-Ashcroft model of 1974~\cite{HamAsh74}.
Such a model is normally  questionable for protons in
an electron gas, as the proton-electron interaction is highly
nonlinear~\cite{JenaSing78,PerrotProton82}. A calculation inclusive
of all non-linear effects is available from  the NPA model,
 and displayed in Fig.~\ref{nei.fig}(a) for a single proton, confirming
that linear-response is accurate at this density. Figure~\ref{BredowH.fig} shows that
the NPA and CHNC agree accurately with each other and with quantum simulations.
While the non-linear CHNC procedure gives  good agreement with the
QMC results of $g_{\rm pp}$, bridge corrections are needed for the form of
the ion-ion pair potential used in the NPA, when excellent agreement is obtained, both
for the positions of the peaks and the peak heights.
The Gibbs-Bogoliubov-LFA criterion determines $\eta=0.475$ for this $\sim$350-times
compressed hydrogen fluid. The $g_{ee}$ and
$g_{e\rm p}$ are insensitive to bridge corrections. However, as expected,
 the accuracy of  $g_{e\rm p}$ depends on the choice of
 $\beta_{e\rm p}\phi_{e\rm p}(r)$. In Fig.~\ref{nei.fig}(b) we see that the
ansatz given by Eq.(35) works well even half-way into the Wigner-Seitz (WS) sphere
of the electron with the WS radius $r_s=0.4473$. Hence the classical map is quite
accurate for equation of state, transport  and other calculations of
 compressed hydrogen in a highly quantum regime.  

The assumptions in Eq.~\ref{Bredow.eq}. that $T_{e\rm p}=\sqrt{T_{ee}T_{ii}}$,
$\phi_{e\rm p}=Z\{1-\exp(-r/\lambda)\}/r$,
 can be avoided if NPA inputs can be used. For instance, in Ref.~\cite{hug02}
the free-electron NPA density $n(r\to 0)$ was used to fix $\lambda_{e\rm p}$.
 A more complete approach is also possible. Thus, if the
free-electron density increment around one H$^+$ ion as  calculated from
the NPA  is $\Delta n_{\rm npa}(r)$, then we
define the $g_{e\rm p}$[one H$^+$]$(q)$ as follows and invert it
by the HNC equation to obtain the effective e-p potential $\beta_{e\rm p}\phi_{e\rm p}$.
\begin{eqnarray}
\label{pot-fromNPA.eq}
h_{e\rm p}[{\rm one}\,H^+](q)&=&\Delta n_{\rm npa}(q,T)/\bar{n}\\
g_{e\rm p}(r)&=&1+h_{e\rm p}(r)=1+\Delta n_{\rm npa}(r)/\bar{n}\\
\label{uei.eqn}
\beta_{ei}\phi_{e\rm p}(r)&=& \mbox{hnc inversion of:}\,g_{e\rm p}(r)
\end{eqnarray}
In Eq.~\ref{uei.eqn} we imply that the
$g_{e\rm p}(r)$ is now interpreted as a classical PDF in a system containing
protons and electrons. It is HNC-inverted to obtain the classical pair-potential
$\beta_{e\rm p}\phi_{e\rm p}(r)$ in a manner analogous to the extraction
of the Pauli potential from $g_{ee}^0(r)$. However, in the regime of high
compressions, the model of Eq.~\ref{Bredow.eq} seems to be sufficient.  

\begin{figure}[t]
\centering
\includegraphics[width=7.0cm, angle=-90]{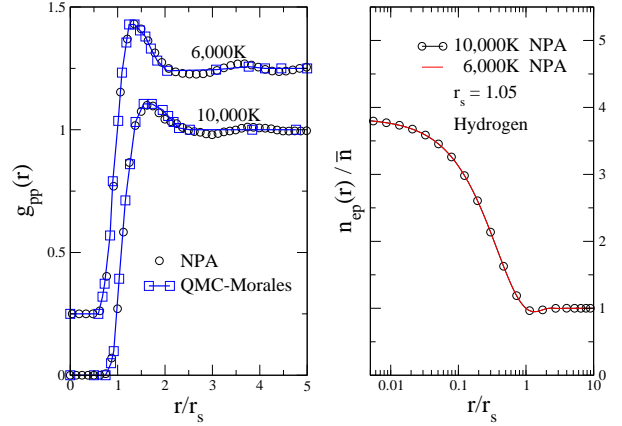}
\caption {(b)The fractional free-electron density $ n_{\rm npa}(r)/\bar{n}$
or $g_{e\rm p}$ at a single proton calculated from the
NPA model at  a proton in a hydrogen plasma at $r_s=1.05$ a.u. This
is effectively unchanged between T=6,000 and 10,000K (b)The  NPA density
displacement can be used to define a pseudopotential for the one-proton system
and  the fluid $g_{\rm pp}$ is calculated using an ion-correlation functional
which reduces to the HNC diagrams, as the Bridge corrections are negligible in
this case. Results agree closely with the coupled electron-ion quantum Monte
Carlo simulations of Morales {\it et a.l}~\cite{Morales2010} which are however limited
 by the size of the simulation box. Only the case $r_s=1.05$ is shown. 
}
\label{npaMorels.fig}
\end{figure}

While the NPA and CHNC calculations agreed with the
QMC results of  Liberatore {\it et al.}  in the linear-response regime, we show
that similar agreement is found in regimes where linear response does
not hold. In Fig.~\ref{npaMorels.fig} we show that the PDFs obtained using our
single-center approaches agree very well with highly computer intensive
 quantum simulations, e.g., those of Morales {\it et al.}~\cite{Morales2010} using
 coupled electron-ion Monte Carlo calculations with 54 protons in the simulation cell.
 The $g_{\rm pp}(r)$ of  such
 quantum simulations
 are limited by the simulation cell size, while the NPA calculations capture the effect
of many Friedel oscillations in the potentials.  Our results imply that
 the ion-ion correlation  functionals used in the single-center NPA method
 to incorporate  many-ion effects are successful. This has also been
verified in many other calculations during past decades, even with respect to
complex fluids like warm dense carbon, silicon etc., where there are covalent
bonding effects as well~\cite{DWP-carb90, CPP-carb18}. A similar
verification is available in the context
of ion-dynamical calculations~\cite{Nadin88,HarbourDSF18}. The NPA and CHNC
methods are not limited by the Born-Oppenheimer approximation, as the static
effects of the electron-nuclear coupling can be incorporated in the
 electron-ion correlation functionals~\cite{Furutani90}.   

The quantum simulation method used by  Morales {\it et al.} is described by them as
``a QMC based ab initio method devised to use QMC electronic energy
 in a Monte Carlo simulation of the ionic degrees of
freedom.... Specifically, the use of twist averaged boundary
 conditions (TABCs)  on the phase of the
electronic wave function, together with recently developed
finite-size correction schemes, allows us to produce
energies that are well converged to the thermodynamic limit
with 100 atoms''. More detailed calculations using CHNC, NPA,
and comparisons of the resulting thermodynamic data 
(calculated from the PDFs using coupling constant integrations)
will be taken up elsewhere.

\section{Systems where all particles are in the quantum regime}
\label{quantum.sec}
If both subsystems, viz., ions and electrons are in the quantum regime, this poses
no additional difficulty for the CHNC method. However, in practice,
such quantum corrections for ions in the liquid state are possible only under extreme
compressions, even for hydrogen, and such compressions are only available
in astrophysical settings. 

On the other hand, in electron-hole systems as found in semiconductor materials,
the quantum nature of both types of particles must be included as the
hole masses (empty states in the valence band) are usually within a factor of
 ten of the electron mass.
Furthermore, if the electrons and holes are confined in two quantum wells
separated by a thin insulating barrier, the spontaneous recombination of
electrons and holes is suppressed. Such systems can be fabricated and are
known as double quantum wells (DQWs) where the electrons  occupy the
the lowest conduction subband in one of the wells, while the holes occupy
the highest valance subband in the other well. They form two interacting
 but spatially separated
2-D electron fluids. Properties of such DQWs in the symmetric case,(i.e., for the
case where the election and hole masses $m_e, m_h$, well widths $w_e,w_h$, layer
dielectric constants and temperatures are equal) have already been studied using
the CHNC method~\cite{lfc-dw19}. However, in typical GaAs-GaAl$_x$As$_{1-x}$ DQWs,   
the electron and hole masses are, typically, 0.067 and 0.335, respectively,
while the material dielectric constant $\varepsilon$ is taken as $\sim12.9$ for
the whole structure since the aluminum alloy content $x$ is small.The barrier
dielectric constant is typically about 12.62.
The effective Bohr radius is given by $a^*_B=\hbar^2\varepsilon/(m_0e^2m^*_s)$ where
$m_0$ is the free-electron mass while $m^*_s$ is the effective mass of the species `s'.
Typical well widths $w$  and barrier widths are 10-30 nm. Given these material
properties, a DQW with equal densities of electrons and holes at a temperature
of even 5K is found to be a two component interacting partially degenerate system
where the $r_s, E_F$ values,
and hence the degeneracies are widely different (see Table 1). In fact,
these DQWs provide excellent laboratory examples of 2-dimensional interacting
warm dense matter. They contain four interacting subsystems as the electrons, and holes
are spin 1/2 fermions for GaAs/GaAlAs  DQWs, but there is no exchange interaction
between particles in separate layers. Such DQWs can be made with two electron layers,
two hole layers, or an electron layer coupled to a hole layer. For more computational
details, see Ref.~\cite{lfc-dw19}.

\begin{table}
\label{eh-rs.tab}
\caption{Characteristic quantities for electrons and holes in a typical GaAs/GaAlAs in
a double quantum well structure. These are examples for 2-dimensional warm dense matter
states where the electron and its positive counter particle behave quantum mechanically.
The material dielectric constant $\varepsilon$=12.9 for both layers separated by
an AlAs barrier of 10 nm width and held at a temperature $T=5K$. Each layer has its
effective Bohr radius and effective Hartree unit. So $r_s, E_F$ etc., are expressed in
the respective effective units. The classical strong coupling parameter
 $\Gamma=1/(r_sT_{cf})$ is given.}
\begin{ruledtabular}
\begin{tabular}{ccc}
 item & electron layer & hole layer \\
\hline
Particle density (cm$^{-2}$)  & 4$\times 10^{11}$ & 4$\times 10^{11}$ \\
Effective mas $m_s^*=m_s/m_0$  & 0.0670            & 0.3350 \\
Layer width (nm)             & 15.00             & 20.00  \\
Effective $r_s$              & 2.768             & 13.84 \\
Effective $E_F$              & 0.1304            & 0.5218$\times 10^{-2}$ \\
degeneracy parameter $T/E_F$   & 0.3015            & 1.508 \\
Classical fluid temp. $T_{cf}/E_F$ &1.966              & 2.188 \\
Inverse de Broglie Length      &1.180              & 0.1807\\
Plasma coupling parameter $\Gamma$ & 1.408         & 6.327 \\
\end{tabular}
\end{ruledtabular}
\end{table}

\begin{figure}[t]
\centering
\includegraphics[width=8cm]{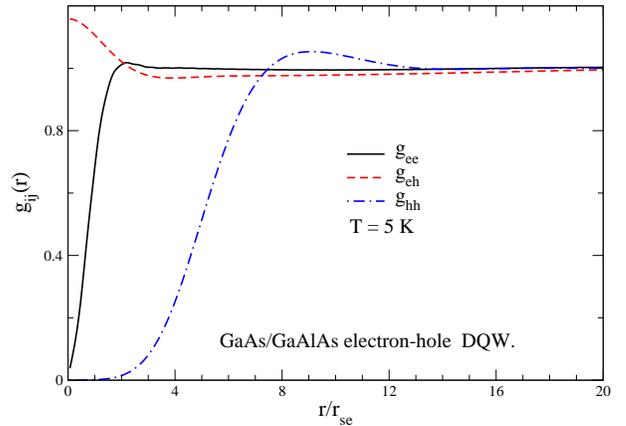}
\caption {The {\it e-e, e-h} and {\it h-h} pair distribution functions $g(r)$ for a
an electron layer interacting with a hole layer in a GaAs/GaAlAs double
quantum well maintained at 5K. The $x$-axis is in units of the electron 
Wigner-Seitz radius $r_{se}=2.768$ in units of the effective electron
atomic unit of length (materials details are given in Table 1).
This is  an example of quasi two-dimensional warm dense matter,
realized at 5K.
}
\label{ehdwq.fig}
\end{figure}

These systems are of great interest in nanostructure physics as the transport properties,
plasmon dispersion, energy relaxation etc., depend on the corresponding structure factors
 and local-field corrections which enter into response functions and effective potentials.
 All such quantities can be extracted from
CHNC calculations~\cite{prl1,lfc-dw19}.
 Currently, the CHNC technique is the only method available for
treating such systems at arbitrary layer degeneracies, spin polarizations,
and arbitrary effective masses at zero to finite temperatures. The
pair-distribution functions for the spin unpolarized $e$-$h$ DQW system described in
Table 1
are displayed  in Fig.~\ref{ehdwq.fig}. Unfortunately, QMC or
other microscopic calculations for such systems are believed to be too prohibitive
at present, and no comparisons are available.

\subsection{The $N$-representability and $v$-representability of CHNC
 densities for electron-ion systems}
The  $N$-representability of the electron-electron $g_{ee}(r)$ obtained from the
CHNC method for electron-ion systems needs to be examined. This too can  be
approached as  in the previous section. It appears that $N$-representability
 is preserved in this case too, where we have merely made
the electrons to interact with the `external potential' of the ions which is, however,
 self-consistently adjusted in the two component problem, with no invoking of the
 Born-Oppenheimer approximation. The BO  corrections come through the electron-ion
 correlation potentials (HNC-like diagrams)
 contained in the HNC equation describing the $g_{e\rm p}(r)$ pair distribution
 function~\cite{Furutani90}.

We may also note that the $v$-representability of the densities generated by CHNC,
or via the NPA can be treated using standard methods since we are dealing entirely
with Coulombic systems and spherical charge densities. For such systems,
Kato's theorem~\cite{Kato57} applies, and
the methods based on spherical densities due to Theophilou, Nagy {\it et al.}
 can be used~\cite{Theoph18,Nagy18}.

\section{Conclusion}
A review of the classical map hypernetted chain procedure, which is a way of side stepping the
construction of a quantum kinetic energy functional for density functional theory, or
doing quantum calculations  is presented.
A proof of the $N$-representability of the classical map, and plausibility arguments
for $N$-representability are given based on agreement of CHNC results with QMC and other
bench mark results.  The application of the CHNC method to general electron-ion
systems is reviewed. Computationally demanding quantum systems like warm dense matter become
numerically very simple and rapid within the CHNC method. The classical map may be used
 without
 the  HNC  procedure, via  classical MD simulations. Thus, within certain limits, entirely
classical calculations which are very rapid and  independent of the number
of particles and the system temperature are  possible for a wide class of
quantum problems.

The author thanks Professors Sam Trickey and Jim Dufty for raising the
question of the  $N$-representability of the CHNC procedure at the CECAM
 workshop at Lausanne, Switzerland, in May 2019.

\end{document}